# A deep learning pipeline for localization, differentiation, and uncertainty estimation of liver lesions using multi-phasic and multi-sequence MRI.

Peng Wang[1*], Yuhsuan Wu[2*], Bolin Lai[3*], Xiao-Yun Zhou[4*], Le Lu[4*], Wendi Liu[1*], Huabang Zhou[1*], Lingyun Huang[2*], Jing Xiao[2*], Adam P. Harrison[4*], Ningyang Jia[1*], Heping Hu[1*]

[1]Eastern Hepatobiliary Surgery Hospital, Shanghai, China
[2]Ping An Technology, Shanghai, China
[3] Georgia Institute of Technology, Atlanta, USA
[4]PAII Inc., Bethesda, Maryland, USA

## Abstract

**Objectives:** to propose a fully-automatic computer-aided diagnosis (CAD) solution for liver lesion characterization, with uncertainty estimation.

**Methods:** we enrolled 400 patients who had either liver resection or a biopsy and was diagnosed with either hepatocellular carcinoma (HCC), intrahepatic cholangiocarcinoma, or secondary metastasis, from 2006 to 2019. Each patient was scanned with T1WI, T2WI, T1WI venous phase (T2WI-V), T1WI arterial phase (T1WI-A), and DWI MRI sequences. We propose a fully-automatic deep CAD pipeline that localizes lesions from 3D MRI studies using key-slice parsing and provides a confidence measure for its diagnoses. We evaluate using five-fold cross validation and compare performance against three radiologists, including a senior hepatology radiologist, a junior hepatology radiologist and an abdominal radiologist.

**Results:** the proposed CAD solution achieves a mean F1 score of 0.62 ± 0.03, outperforming the abdominal radiologist (0.47), matching the junior hepatology radiologist (0.61), and underperforming the senior hepatology radiologist (0.68). The CAD system can informatively assess its diagnostic confidence, i.e., when only evaluating on the 70% most confident cases the mean f1 score and sensitivity at 80% specificity for HCC vs. others are boosted from 0.62 to 0.71 and 0.84 to 0.92, respectively.

**Conclusion:** the proposed fully-automatic CAD solution can provide good diagnostic performance with informative confidence assessments in finding and discriminating liver lesions from MRI studies.

# Introduction

Liver cancer is one of the most common cancers worldwide [1]. Among malignant liver lesions, there are several types of masses, including hepatocellular carcinoma (HCC), intrahepatic cholangiocarcinoma (ICC), metastasis from other primary sites, and other rare tumors. Differentiating liver lesions is critical in clinical practice since the treatment options and prognosis depend on the diagnosis [2, 3]. In particular, HCC is the most common primary liver cancer [4] and differentiating HCC lesions from other types is a major focus in clinical care, e.g., the Liver Imaging Reporting and Data System (LI-RADS) protocol is focused on this very problem [5]. Non-invasive imaging techniques are crucial in clinical workflows, with multi-phasic MRI providing the highest performance [6]. In routine clinical diagnosis, the first step is to localize the tumor location and then distinguish the tumor type. This work is not only workload-heavy, but performance is also dependent
on the clinician's experience. Additionally, specificity in HCC diagnosis for small nodules, even with MRI, can be low [6]. Machine learning computer-aided diagnosis (CAD) solutions to provide decision support could be beneficial for liver lesion characterization.
The utility of using machine learning, specifically convolutional neural networks (CNNs), to diagnose liver lesions using MRI has been previously verified [7, 8]. However, prior studies rely on manually drawn regions of interest (ROIs) around lesions [7, 8] or even complete tumor delineations [9], which in a deployment scenario would still rely on radiologists' assistance that is laborious and a potential source of inter-rater variability. CAD systems that classify lesions from raw MRI images have also been reported, but these are likely vulnerable to overfitting [10]. Besides differentiating lesion types, it is also important for a CAD system to provide a confidence of any prediction. In clinical practice, the confidence of any diagnosis is an important factor in treatment selection [11]. Physicians may request second opinions or seek additional clinical material if they are unsure of their diagnosis. Similarly, CAD systems should also output a confidence on a case-by-case basis [12] and, ideally, higher confidences would correlate with higher performance.
We propose a fully automatic CNN-based CAD system for multi-phasic and multi-sequence MR that localizes liver lesions, differentiates their type, and provides a predictive confidence uncertainty. As proof-of-concept, we validate our system on differentiating between HCC, ICC, and metastasized lesions on retrospectively collected and histopathology-confirmed MRI studies, comparing our system's performance to three board-certified radiologists.

# Materials and Methods

## Data Collection and Imaging Acquisition
MRI studies were retrospectively collected from the picture archiving and communication system (PACS) of the Eastern Hepatobiliary Surgery Hospital in Shanghai, China. The selection criteria were any patient who had a liver surgical resection or biopsy in the period between 2006 and 2019 and was diagnosed with an HCC, ICC, or secondary metastasis lesion. All metastasis patients were diagnosed with an adenocarcinoma. Enrolled patients must have available an MRI study within one month prior to the histopathology procedure. All studies were collected from 3-T MR GE Discovery scanners and the studies comprised T1WI, T2WI, T1WI venous phase (T2WI-V), T1WI arterial phase (T1WI-A), and DWI (b-values = 800 - 1000 s/mm^-2) sequences. Because delay phase scans were not universal, we did not incorporate them. For multi-phasic T1WI, three scans were obtained with gadolinium-based contrast agents at a 30ml dose, which included a pre-contrast phase, arterial phase (~ 20 s post-injection), and portal venous phase (~ 50 s post-injection). Additional acquisition parameters are listed in Table 1. This resulted in 926 multi-phasic MRI studies, from which we selected all with ICC and metastasis diagnoses (83 and 110, respectively) and randomly selected 207 HCC patients to total 400 studies. Figure 1 depicts the patient selection process and Table 2 outlines the clinicopathological details of the dataset. Under the supervision of and verified by a hepatologist with more than 10 years of experience, each malignant lesion was annotated on each slice using RECIST marks [13]. When lesions were too numerous to individually annotate, a mark was drawn across each lesion cluster. From these, a set of 2D bounding boxes were generated for each study, whose number totaled 13247.

## Deep Learning Model for Key ROI Localization

To localize ROIs we use an ROI regression pipeline inspired by key slice parsing [14]. In short, we train five separate 2D deep learning detectors, one for each MRI sequence type. As detectors we use the CenterNet framework [15] with a DLA34 [16] backbone. Because of the importance of T2WI in discovering lesions [6], we include use it as a second channel (except for the T2WI detector, which we just train on T2WI alone). During inference we remove any poor-quality ROIs whose confidence is lower than 0.5 [14]. As a result, occasionally the localizer will report that it is unable to localize any ROIs. We then regress the most likely lesion ROI for each slice using a voting scheme, whereby each pixel in the slice receives a vote if any bounding box from the five detectors overlaps with it. The pixel with the most votes corresponds to the regressed ROI location. The size and confidence of the final ROI correspond are the corresponding average values from all overlapping bounding boxes. Finally, we only pass on the most confidence ROIs for each study, using the criterion (top 48%) reported by Lai et al. [14]. A listing of ROI localization hyperparameters and more details can be found in the Supplementary Material.

## Deep Learning Model for Lesion Characterization

After localization, we use a deep 2D Densent121 [17] CNN to classify ROIs into three types: HCC, ICC, or metastasis. All five MRI sequences are concatenated into five channels as input to the network. During inference, we input the ROIs produced by the detection model to obtain ROI-wise pseudo-probabilities. These are then averaged to produce the study-wise pseudo-probability. For categorical outputs, the lesion type with the highest pseudo-probability is predicted.

It is well understood that the magnitude of pseudo-probability outputs from deep CNNs are not good indicators as to their reliability [12]. Recent studies have demonstrated that an informative measure of uncertainty can be obtained by applying dropout during inference to generate a distribution of outputs. The distribution's dispersion can then capture model uncertainty [18, 19]. We follow this procedure to produce an ROI-wise
uncertainty by applying dropout before the final fully-connected layer, setting the dropout probability to 0.2, forward passing each ROI 100 times, and finally calculating the variance of the predictions. To produce a study-wise uncertainty, we compute the
average variance across all ROI-wise uncertainties. We then convert the uncertainty to a confidence by simply subtracting its value from 1. A listing of the classifier hyperparameters and more details can be found in the Supplementary Material.

### Reader Study

To compare against human readers, we measured the diagnosis performance of three board certified radiologists (Radiologist A with 20 years' experience in hepatic imaging, Radiologist B with 8-years' experience in hepatic imaging, and Radiologist C with 15 years' experience in general abdominal imaging) on the same 400 MRI studies. Readers were blinded to the histopathological diagnoses and any other clinical data. The same five sequences used by the CAD system were presented and readers were asked to differentiate cases into the three sub-types. Readers were given the original sequences in DICOM format and were free to use their viewer of choice. Additionally, readers were also asked to rate the confidence of each diagnosis using an ordinal scale ranging from 1 to 5, corresponding to "Not Confident at All", "A Little Confident", "Reasonably Confident", "Confident", and "Extremely Confident", respectively.

### Statistical Analysis and Evaluation Metrics

All 400 patients were split using five-fold cross validation, with 80%, 10%, and 20% used for training, validation, and testing, respectively, in each fold. Data splitting was executed on HCC, ICC, and metastasis independently to avoid imbalanced distributions. We report average performance metrics across folds along with their standard deviation, providing a measure of the variation caused by differences in training and testing sets. To evaluate categorical predictions, we measure the accuracy and the sensitivity, specificity, and F1 score for each lesion type. Should the CAD system fail to localize any ROIs for a study, that case is treated as a false negative for the ground-truth lesion type in
question. Given the clinical focus on HCC [5, 6], we also perform localization ROC (LROC) analysis [20] on the HCC vs. others performance of the CAD system. LROC analysis is like ROC analysis, but any HCC studies where an ROI could not be found are treated like false negatives, meaning the maximum sensitivity of the CAD system may not reach 100%. We avoid constructing ROC curves for the readers using their self-reported confidence scores because there is convincing analyses that such scores do not map well to actual clinical operating points [21, 22] and can violate the assumptions of ROC
analysis [23]. Thus, we simply examine reader sensitivity and specificity for discriminating HCC vs. others without considering their diagnostic confidence.
To properly explore the relationship between diagnostic confidence and performance for both the CAD system and the readers, we calculate our metrics of interest across different confidence cutoffs. To be specific, we remove cases with a study-wise confidence lower than a particular threshold and measure performance on the remaining patients. We expect performance to increase as non-confident cases are removed, but at the cost of leaving some patients undiagnosed. For categorical predictions, we plot accuracy and mean F1 scores across different confidence levels, along with the corresponding number of cases that were retained. For HCC vs others performance, we perform LROC analysis on a subset of definitive diagnoses (70% of patients kept).

## Results

### Lesion type characterization

Table 3 reports the accuracy, sensitivity, specificity, and F1 scores of our model and the three radiologists on HCC, ICC, and metastasis characterization. Out of 400 studies, the CAD system failed to localize any ROIs for 8 studies (HCC=6, ICC=1, metastasis=1). Our model's mean F1 score of 62% is comparable to radiologist B (61%), outperforms radiologist C (49%), and is lower than the more senior radiologist A (68%). Both the model and the readers perform best in discriminating HCC, worst in discriminating ICC, and somewhere in between in discriminating metastasized lesions. Figure 3(a) depicts the LROC curve of the CAD system with the operating points of each radiologist also rendered. Our model performs comparably to radiologists A and B while outperforming radiologist C. Figure 3(b) and (c) depict LROC curves for patients whose largest lesion was >2cm (n=331) and <=2cm (n=69), respectively. Interestingly, for small lesions radiologist B outperformed radiologist A, despite the latter's better overall performance. For the small-tumor subset, our CAD system underperforms both radiologists A and B, but radiologist B overlaps with the CAD system's across-fold variance. Overall, from Table 3 and Figure 3, it can be concluded that, on our dataset, the CAD system performs slightly worse than a radiologist with 20 years' experience in hepatic imaging (radiologist A), performs comparably to a more junior hepatic radiologist (radiologist B), and better than a radiologist not specialized in hepatic imaging (radiologist C). Figure 4 provides some representative examples. As can be seen, there can be reader disagreement.

### Diagnostic Confidence Study

Next, we evaluated the impact of the uncertainty estimates for both our CAD system and that of the readers. For the readers we focus on radiologists A and B, who specialized in the liver and who tended to perform better. The results for radiologist C results can be found in the supplementary (Fig. S1). Figure 5 depicts the accuracy and mean F1 scores of the CAD system and readers across different confidence levels. Note that for the CAD system, the confidence is specified using a continuous value from 0.7 to 0.95, while for the two radiologists, the confidence is specified using discrete values from 1 to 5. Both the CAD system and the two physicians were able to informatively assess certainty. When keeping the 70% most definitive predictions, the CAD system boosts the accuracy from 68% to 79% and the mean F1 score from 62% to 71%. At the same operating point of keeping 70% of the predictions, radiologist A (radiologist B) boosts the accuracy from 73% to 81% (66% to 70%) and the mean F1 score from 68% to 76% (61% to 67%). Figure 6 depicts the HCC-vs.-others

LROC of the CAD system when keeping the 70% most certain predictions. The overall AUC boost is only marginal (from 89% to 91%) but from examining the LROC curve performance gains are high at high-specificity operating points (which are the clinically useful regions). For instance, at the sensitivity at 80% specificity increases from 84% to 92%.

## Discussion

Characterizing malignant liver lesions is important for oncological diagnosis and treatment [2, 3]. Pathological diagnosis is the gold standard, but it is invasive. Observation of MRI images, along with other clinical information, can provide a non-invasive diagnosis. Yet, the inter-rater reliability when using the LI-RADS protocol has been reported to be only moderate to good [24, 25]. Moreover, the LI-RADS protocol is not universally applied, due to its complexity and other issues [26, 27]. Thus, inter-rater reliability in many clinics or locales may not match reported LI-RADS numbers. This study demonstrated that a fully automated deep-learning CAD system can provide good diagnostic performance (HCC vs. others AUC=89%) in differentiating lesions from MRI on a large dataset of 400 patients. The CAD system's mean F1 score of 0.62 is comparable to two radiologists specialized in hepatic imaging (mean F1 of 0.61 and 0.68) and outperformed a radiologist specialized in general abdominal imaging (mean F1 of 0.47). Moreover, we demonstrated that our CAD system can informatively assess its diagnostic confidence.

Previously reported CAD systems are typically not fully automatic. They may expect radiologists to manually indicate an ROI [7, 8] or, more laboriously, to delineate a lesion mask [9]. Apart from raising barriers to clinical adoption, manual localizations also introduce inter-user variations, which are undesirable confounders. One exception is Trivizakis et al., who directly classified whole 3D MRI scans without any localization [10]. Given the overfitting tendencies of deep networks, such an approach could be susceptible to fitting to imaging features and regions not relevant to liver lesion differentiation. In contrast, we report a localization + classification CAD pipeline to reliably localize and then classify lesion ROIs. We evaluate our approach on a dataset of 400 MRI studies that are sampled from a clinical population with minimal selection criteria (Figure 1) and that is larger than prior test sets, which range from 23-200 [7–10]. Importantly, we perform cross validation, which can help illuminate how stable the CAD system is across choices of training and test sets. Our fully automatic pipeline can achieve good diagnostic performance that compares well to clinical experts. For instance, our model's F1 score of HCC vs. others is 0.82, which is comparable to a radiologist with 8 years' specialization in hepatic imaging (HCC vs others F1 of 0.82) and compares well to reported scores of 0.82 in the literature [6].

Another distinct capability is that our CAD system produces diagnostic-confidence assessments. Separating definitive diagnoses from non-definitive ones can help identify when to seek additional opinions or when to order further tests, e.g., an invasive biopsy procedure [12]. Several studies suggest that diagnostic confidence is not always well calibrated For human readers [28, 29]. Because they can be calibrated, in principle CAD systems have a comparative advantage. However, deep learning pseudo-probabilities are well known to be unreliable gauges as to confidence [12]. Monte Carlo dropout is a recently proposed enhancement for confidence assessment [18, 19]. As far as we know, there have been no investigations on the relationship between a proper deep learning confidence measure and diagnostic performance for lesion characterization. As the results demonstrate, our CAD system can separate definitive and non-definitive diagnoses. For instance, when limiting evaluation to the 70% most confident predictions, the sensitivity at 80% specificity for HCC vs. others is boosted from 84% to 92% (Fig. 6), with commensurate improvements in its categorical accuracy and mean F1 scores (Fig. 5). In fact, when only considering the 70% most confident diagnoses, the CAD system's mean F1 score of 0.71 is higher than all other radiologists if they assess all patients. This suggests that the system could even be selectively applied depending on performance demands, indicating when it "does not know" otherwise.

There are several limitations in our study. First, our dataset is class-imbalanced and performance in the ICC-vs-others and metastasis-vs-others metrics (Table 3) suffers as a result. However, human readers also performed relatively worse for these lesion types. Another limitation is that benign lesions were not included in the analysis. Additionally, all studies were pathology proven but this restriction does apply a selection bias. Future efforts at data collection are necessary. Small lesions (<2cm) challenged our CAD system, where a larger gap opened between it and radiologists' performance (Figure 4). Reducing this gap is crucial. In terms of the reader study, it did not reflect actual clinical conditions, as the radiologists only had access to the images themselves without any ancillary information. Actual clinical performance would likely be higher. Nonetheless, this allowed our CAD system to be compared against radiologist interpretation when presented with the same information. Finally, expanding the work to include multiple institutions and multiple scanners is necessary to verify the findings from this feasibility study.

In conclusion, we presented a feasibility study of a liver lesion differentiation pipeline for MRI. Unlike prior work, we report a fully automatic approach that requires no manual localizations. We evaluated on 400 pathology-proven multi-phasic MRI studies using cross validation. The CAD system can match a hepatic radiologist with 8 years' experience, slightly underperform a more senior hepatic radiologist with 20 years' experience, and outperform an abdominal radiologist with 15 years' experience. Moreover, our CAD system produces informative assessments of its diagnostic confidence, providing an additional key piece of information to aid clinical decision making. This study demonstrates that our CAD approach could be a valuable supporting tool.

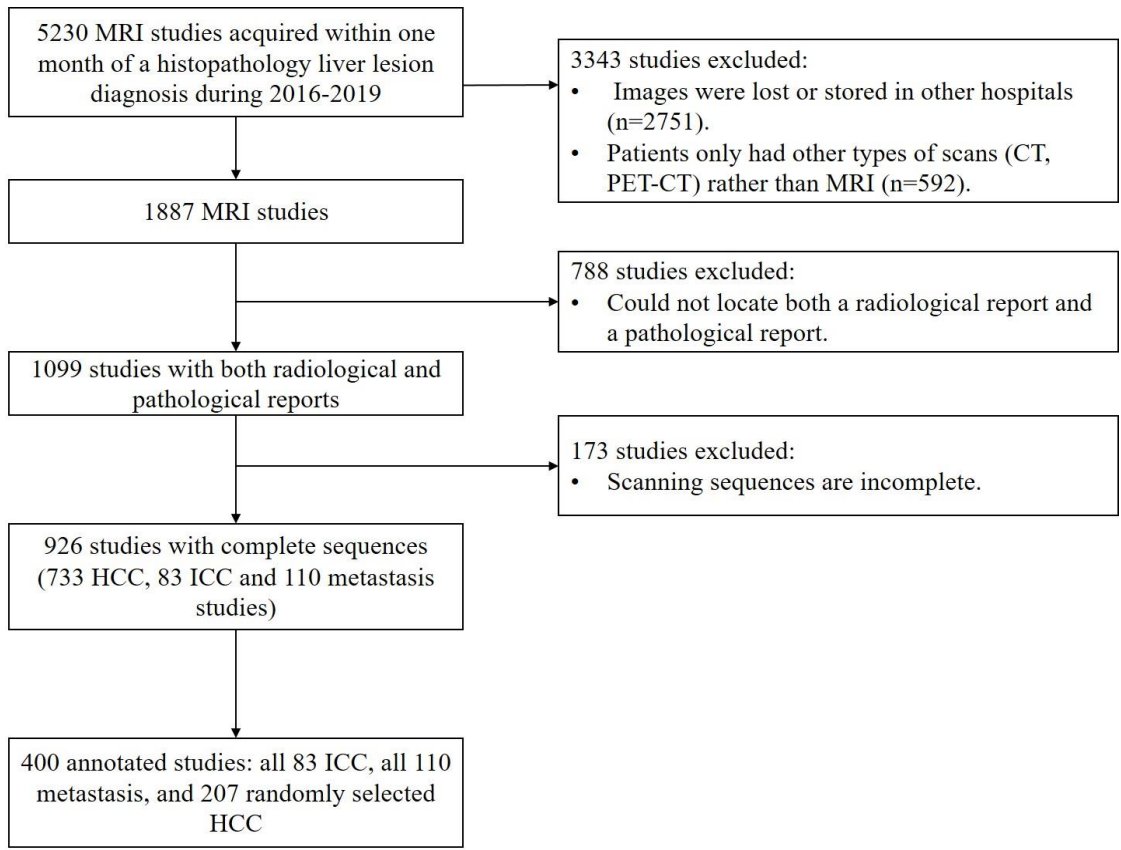

**Fig. 1** Data collection and annotation flowchart.

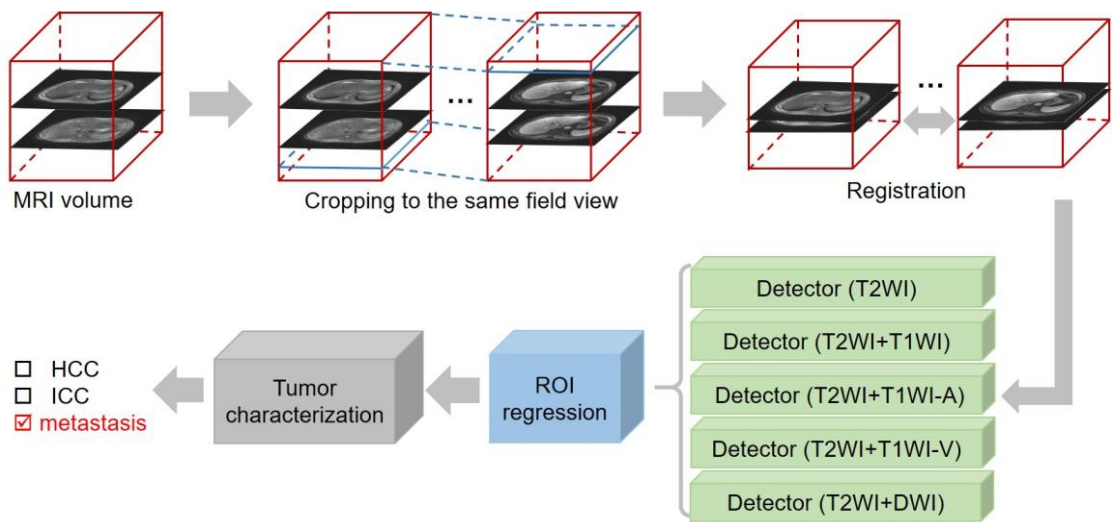

**Fig. 2** Illustration of the deep learning pipeline for tumor localization and differentiation.

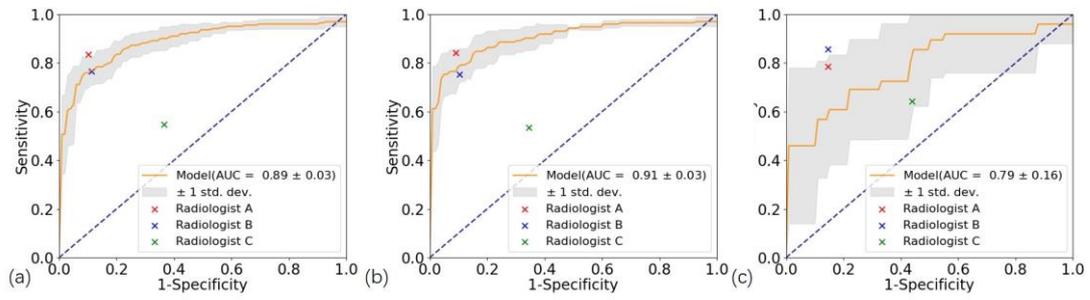

**Fig. 3** The HCC vs. others LROC curves for our CAD system (orange line). Note, because the CAD system reported that it could not localize lesions in some studies (n=8), the maximum sensitivity is less than 100%. Also rendered are the operating points for all three radiologist readers (points). (a) depicts performance on the overall dataset, while (b) and (c) show performance on patients whose largest tumor is >2cm (n=331) and <=2cm (n=69), respectively.

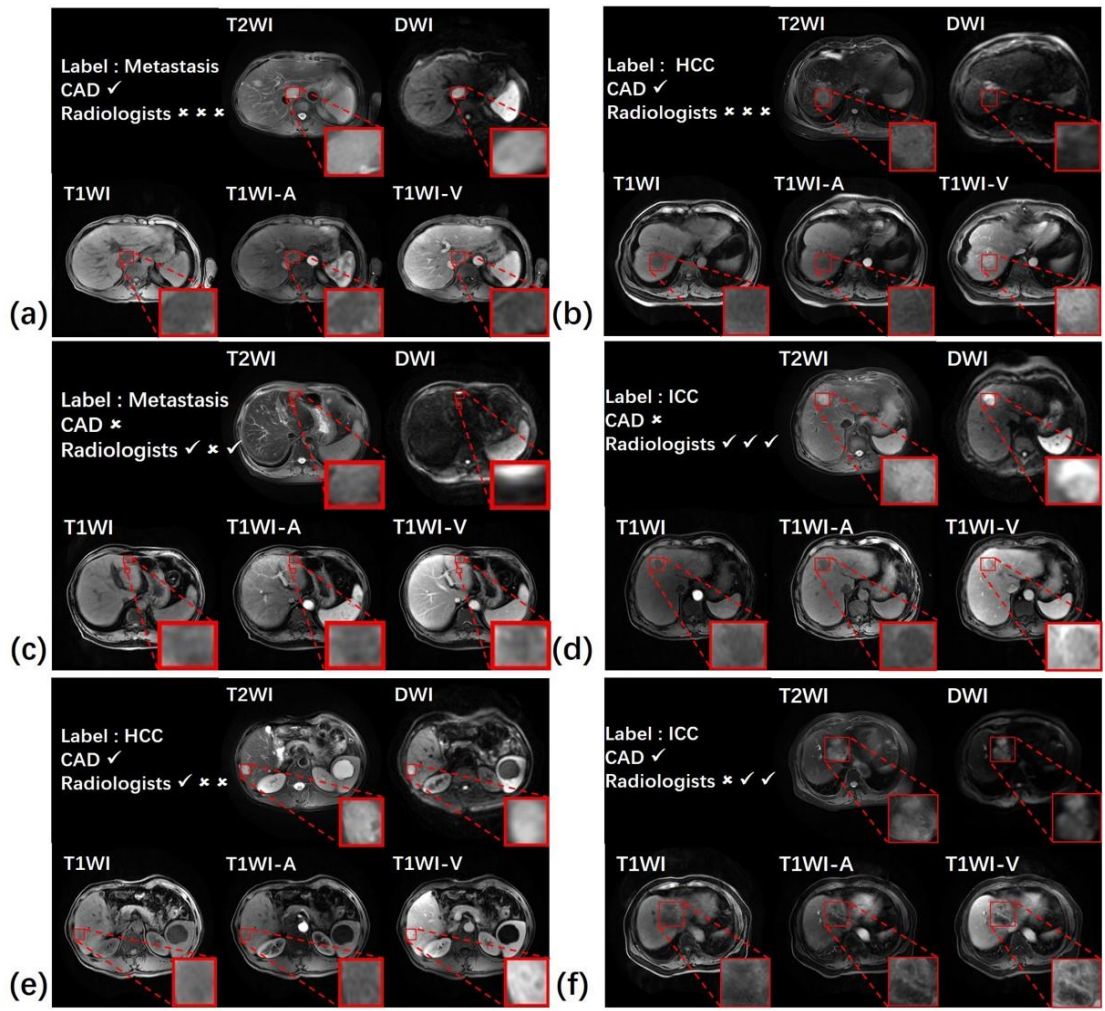

**Fig. 4** Visualization on five sequences of the six patients. Lesion sub-types and the performance of three readers and our CAD system are shown on the top left corner. The ticks and crosses denote correct and incorrect classification results, respectively. (a) and (b) show examples where the CAD system made a successful diagnosis, but all three readers did not. (c) and (d) show failure cases. As can be seen, readers also disagreed on these cases. (e) and (f) show success cases where only some of the radiologists made a successful diagnosis.

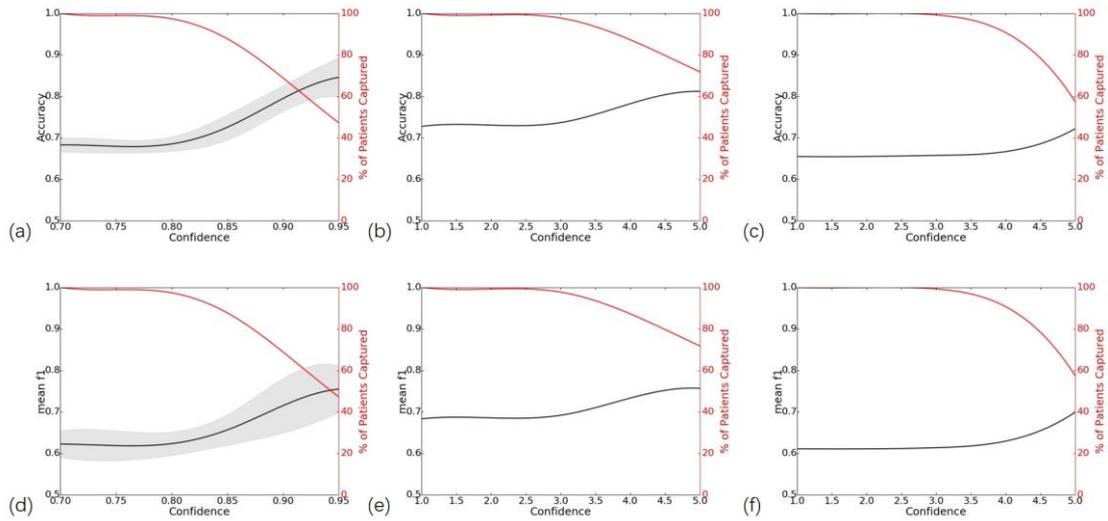

**Fig. 5** Overall diagnostic performance (black line) and corresponding percentage of captured patients (red line) at different confidence thresholds. (a)-(c) show the accuracy of the CAD system, radiologist A and radiologist B, respectively. (d)-(f) show the mean F1 score of the CAD system, radiologist A and radiologist B, respectively.

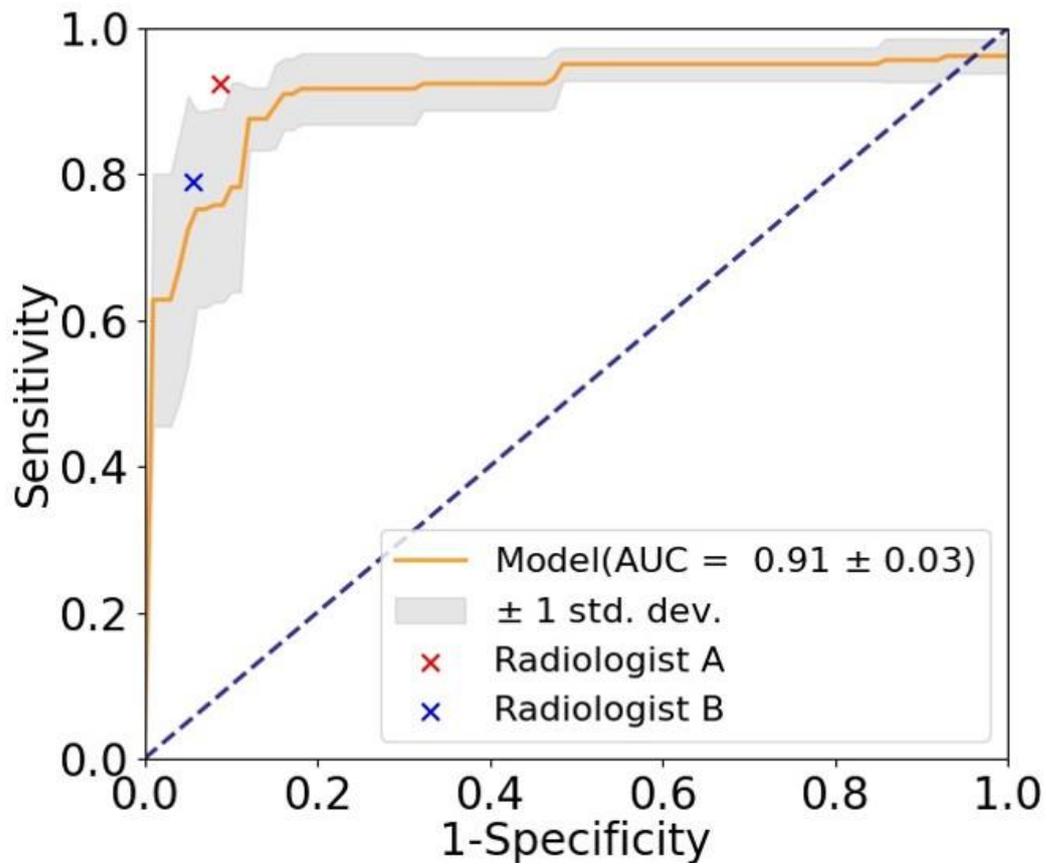

**Fig. 6** The HCC vs. others LROC curve for our CAD system (orange line) and two radiologist readers (points) when keeping the 70% most confidence predictions.

Table 1: MRI Acquisition parameters used for the MRI dataset.

| Sequence | TR (ms) | TE (ms) | Acquisition time (s) | Flip angle | Matrix | In-plane Resolution(mm) | Slice thickness (mm) |
|---|---|---|---|---|---|---|---|
| T2WI | 2222-16363 | 68.1-77.2 | 7.6-21.1 | 110° | 320 x 320 | 0.6-0.8 | 6.0-8.0 |
| Multi-phasic T1WI | 3.6-3.7 | 1.7 | 8.0-21.1 | 12° | 260 x 260 | 0.6-0.8 | 4.4-7.0 |
| DWI | 2849-17142 | 44.6-55.1 | 8.0-21.9 | 90° | 128 x 128 | 1.2-1.7 | 6.0-10.0 |

Table 2: Clinicopathological details of the MRI study dataset. Size metrics are for the largest lesion in each patient. When lesions were too numerous to individually delineate with bounding boxes, lesions clusters were annotated instead. Thus, size metrics include both lesion and lesion clusters.

|  | Total (n = 400) | HCC (n = 207) | ICC (n = 83) | Meta (n = 110) |
|---|---|---|---|---|
| **Mean Age ±SD** | 56±11 | 54±10 | 58±11 | 57±11 |
| **Sex** | | | | |
| **Men (%)** | 292 (73%) | 172 (83%) | 53 (64%) | 67 (61%) |
| **Women (%)** | 108 (27%) | 35 (17%) | 30 (36%) | 43 (39%) |
| **Lesion Characteristics** | | | | |
| **Median Size (cm)** | 3.94 | 3.97 | 5.43 | 3.11 |
| **Mean Size ±SD (cm)** | 5.14±3.74 | 5.08±3.01 | 6.31±4.25 | 4.39±3.68 |
| **>2cm (%)** | 331 (83%) | 179 (86%) | 77 (93%) | 75 (68%) |
| **<= 2cm (%)** | 69 (17%) | 28 (14%) | 6 (7%) | 35 (32%) |
| **Liver Cirrhosis** | | | | |
| **Yes (%)** | 160 (40%) | 127 (61%) | 25 (30%) | 8 (7%) |
| **No (%)** | 240 (60%) | 80 (39%) | 58 (70%) | 102 (93%) |

Table 3. Categorical performance metrics of the three radiologist readers and our CAD system. CAD metrics correspond to the mean value across the five cross validation folds with the standard deviation indicated in parentheses. Note, the 8 studies where the CAD system failed to localize any lesions are treated as misdiagnoses for accuracy and false negatives for the lesion-type specific metrics.

|  |  | Radiologist A | Radiologist B | Radiologist C | CAD System |
|---|---|---|---|---|---|
|  | Accuracy | 0.73 | 0.66 | 0.49 | 0.68 ± 0.02 |
|  | Mean F1 score | 0.68 | 0.61 | 0.47 | 0.62 ± 0.03 |
| HCC vs. others | Sensitivity | 0.84 | 0.77 | 0.55 | 0.82 ± 0.06 |
| HCC vs. others | Specificity | 0.90 | 0.89 | 0.63 | 0.80 ± 0.09 |
| HCC vs. others | F1 score | 0.86 | 0.82 | 0.58 | 0.82 ± 0.04 |
| ICC vs. others | Sensitivity | 0.63 | 0.73 | 0.50 | 0.42 ± 0.14 |
| ICC vs. others | Specificity | 0.79 | 0.67 | 0.61 | 0.89 ± 0.03 |
| ICC vs. others | F1 score | 0.52 | 0.49 | 0.34 | 0.44 ± 0.11 |
| Metastasis vs. others | Sensitivity | 0.60 | 0.39 | 0.36 | 0.62 ± 0.09 |
| Metastasis vs. others | Specificity | 0.92 | 0.97 | 0.96 | 0.83 ± 0.06 |
| Metastasis vs. others | F1 score | 0.66 | 0.53 | 0.50 | 0.61 ± 0.05 |